\documentclass[runningheads]{llncs}
\usepackage{url}
\usepackage{graphicx}
\usepackage{amsmath,amssymb,amsfonts}
\begin{document}
\title{Belief-Desire-Intention (BDI) Multi-agent System for Cloud Marketplace Negotiation}
\titlerunning{BDI Multi-agent System for Cloud Marketplace Negotiation}
%
\author{Saurabh Deochake}
\authorrunning{Saurabh Deochake}
%
\institute{Twitter, Inc., San Francisco CA 94103, USA\\ 
\email{sdeochake@twitter.com}}
\maketitle              
\begin{abstract}
With the evolution of cloud computing, there has been a rise of large enterprises extending their infrastructure and workloads into the public cloud. This paper proposes a full-fledged framework for a Belief-Desire-Intention (BDI) multi-agent-based cloud marketplace system for cloud resources. Each party in the cloud marketplace system supports a BDI agent for autonomous decision making and negotiation to facilitate automated buying and selling of resources. Additionally, multiple BDI agents from an enterprise competing for the same cloud resource can consult with each other via Master Negotiation Clearing House to minimize the overall cost function for the enterprise while negotiating for a cloud resource. The cloud marketplace system is further augmented with assignments of behavior norm and reputation index to the agents to facilitate trust among them. 

\keywords{Intelligent Systems, Multi-agent Systems, Cloud Computing, Automated Negotiation, BDI Agents}
\end{abstract}
\section{Introduction}
With the advent of rapid growth in public cloud computing in recent years, enterprises that traditionally operated their own data centers have started their journey of using the public cloud. Public cloud platforms offer Infrastructure as a Service (IaaS) while delivering the scalability, efficiency and elasticity to operate services that can be much better than running them in data centers. Using the cloud, an enterprise can easily scale their infrastructure using on-demand resource utilization via a pay-as-you-go pricing model. An important benefit of having on-demand resource availability is that the enterprise does not have to purchase expensive hardware in advance, thereby reducing their capital expenditure \cite{cloud_cost}. Currently, all major public cloud platforms offer marketplace systems that sell pre-built products that are implemented according to the infrastructure specifications specific to that cloud provider. For example, Amazon Web Services (AWS) has implemented the Spot Instance Marketplace where if a buyer's bid matches or exceeds the marketplace price, that Spot Instance starts running \cite{ec2}. While these marketplaces exist in closed environments specific to those cloud providers, there is a need for comprehensive efforts around building an inclusive and cross-platform cloud marketplace. Moreover, the bidding process to acquire a resource on the cloud marketplace requires substantial human intervention in sending the request for quotes, finding the best price for the resource, predict the capital expenditure to acquire the resource, and finally negotiate with the seller. \\
\indent Negotiation is a mechanism in which two or more parties reach a consensus based on a joint future venture. Here, each of these parties may have an intelligent agent negotiating on behalf of them with opponent agents. An agent can also be assigned to an infrastructure service for an enterprise, thereby enabling multi-agent system in a scenario where multiple services are autonomously expanding the infrastructure footprint by buying necessary cloud resources from the marketplace. In order to pursue the best deal, these agents must negotiate with multiple sellers from various cloud providers on the cloud marketplace to select the best possible outcome. Therefore, the transactions that involve offers and counter-offers assume alliances that may be ad hoc or permanent. In the case of cloud marketplace system, the negotiation agenda may include a single resource issue such as price of virtual machine or multiple issues for a resource like the type of CPUs, amount of memory, and persistent disk size for a virtual machine. Therefore, there is a need for a multi-agent cloud marketplace system that supports the automated negotiation and decision-making that assists enterprises in autonomously expanding their infrastructure footprint into the cloud.\\
\indent This paper proposes a comprehensive architecture for a cloud marketplace that is based on BDI multi-agent system. This paper is structured as follows: Section \ref{sec:related_work} discusses the previous work, section \ref{sec:system} showcases the cloud marketplace architecture, section \ref{sec:negotiation} presents the approach for optimizing the overall cost function for the self-competing agents from the same enterprise, section \ref{sec:future_work} discusses the future work, and finally, section \ref{sec:conclusion} concludes this paper. 

\section{Related Work}\label{sec:related_work} 
BDI software model of programming the agents has been a part of multiple research studies, especially because BDI agents exhibit philosophical attitudes like beliefs, desires and intentions that can be applied in real life \cite{nego_prospects}. Beliefs are agent's current understanding of the world. Desires are the objectives that the agent may plan to execute based on the current knowledge about the surrounding world. Intentions represent the desires that agent has committed to execute. Although a BDI agent is a robust model for a multitude of real world applications, there are a few limitations on the traditional BDI agents such as lack of learning \cite{bdi_learning}, future planning and multi-agent system integration. Since the BDI agents' intentions are desires that the agent has almost committed to execute, traditional BDI agents lack a robust mechanism to learn from the past behavior and look-ahead planning. Finally, the traditional BDI model does not describe the mechanisms for integration of a BDI agent into a multi-agent system \cite{bdi_mas}. This paper proposes an architecture that constitutes a BDI multi-agent system for a cloud marketplace application. \\
\indent Previous work like \cite{nego_bilateral} showcases a way to achieve bilateral multi-issue negotiation where two agents may choose to negotiate on the issues concurrently. While this can be a viable option to perform concurrent negotiation, it also leads to over-negotiation between the agents that continues over an extended period of time costing more time and resources. A novel approach of utilizing beliefs, goals and plans for agent negotiation was introduced in \cite{nego_bgp}. The BGP-based multi-strategy negotiation allows an agent to select the best suitable negotiation strategy for an occasion. Moreover, the agents participating in the negotiation may send the offers on the issues that may be mutually exclusive in terms of importance to each agent. In this situation, the negotiation may not conclude in a reasonable amount of time. A previously agreed upon list of issues and a predetermined time threshold will assist both agents to conclude their negotiation within a finite time on a mutually inclusive set of issues. The work presented in \cite{nego_time_behavior} provides a guidance on time and behavior dependent negotiation strategy where the agents select issues with the least weight of importance associated with them to concede during the negotiation. \\
\indent Traditionally, the applications of autonomous agents in negotiation have been in the area of e-commerce. The author's previous work, for example, relies on the techniques of the game theory principle to select a predefined negotiation strategy to reach the consensus \cite{nego_henri,nego_mainwave}. The work primarily focuses on a win-win strategy where the agents try to reconcile their utilities in order to reach the solution. When it comes to the application of agent negotiation in cloud computing use cases, to the author's best knowledge, the most of the work focuses on cloud computing service level agreements (SLAs) between a cloud service user and cloud service provider \cite{nego_cloud_sla_1}. While there is an attempt to study the multi-cloud negotiation system via a cloud agency-based approach, like mOSAIC \cite{nego_mosaic}, it does not provide an explicit guidance on the concurrent multi-party, multi-issue negotiation methodology. For an enterprise to react to the increasing load on their infrastructure, it may require its infrastructure to expand vertically. Since the cloud resources are likely available on a pay-as-you-go basis, the cloud becomes a viable alternative for the elastic expansion of the infrastructure. The author believes that the employing the BDI agents would empower the autonomous expansion and contraction of the infrastructure by negotiating the resources over the cloud marketplace system. The author's previous work \cite{nego_agent_hybrid_cloud} uses a time-resource deadline-based negotiation strategy by using autonomous buyer and seller agents to reach an agreement over various issues. However, the model presented in the paper assigned only one agent for an enterprise to negotiate with the opponent cloud provider agent. It did not provide any flexibility for an individual infrastructure service to scale vertically. This paper tries to solve that limitation by assigning a BDI agent for every infrastructure service, thereby constituting a multi-agent system and providing flexibility for a service in an enterprise.

\section{System Architecture}\label{sec:system}
\begin{figure}[t]
  \centering\includegraphics[width=0.9\linewidth]{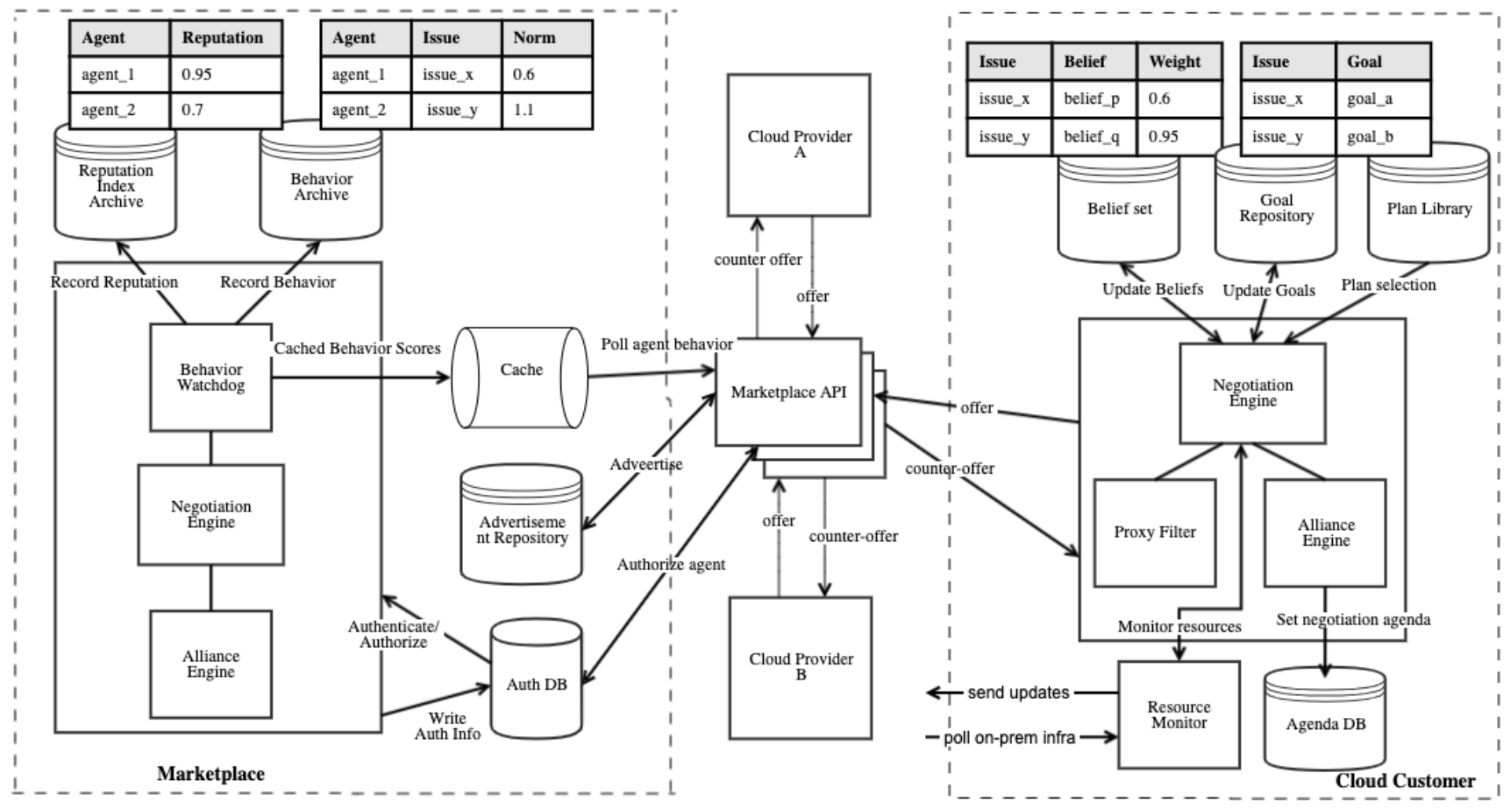}
  \caption{A Bird's Eye View of The Cloud Marketplace System}
  \label{fig:overview}
\end{figure}

The components of the Cloud Marketplace architecture described below are designed in the author's previous work \cite{nego_agent_hybrid_cloud}. Figure \ref{fig:overview} showcases the bird's eye view of the system architecture from the previous work. 

\subsection{Cloud Marketplace}
The Cloud Marketplace is a highly available distributed system that implements all crucial components of a marketplace facilitating the negotiation among agents. While the standard methods are used to build distributed systems, the components of the marketplace are made available to the agents via RESTful API interfaces. The following subsections briefly discuss major components of the cloud marketplace system. 

\noindent \textbf{Advertisement Repository}: 
Advertisement Repository is a distributed data-base system that stores the advertisements sent by agents that operate in the cloud marketplace. These advertisements contain information related to the agents and the product like the agent ID, product ID, and a subset of features or issues for the product. The agents can then query the Advertisement Repository to find the suitable opponent agent as per the \texttt{requestForQuote} message to start the negotiation rounds.

\noindent \textbf{Alliance Engine}: 
Alliance Engine also acts as a match-maker for the agents by polling the Advertisement Repository. When an agent sends an \texttt{RFQ} message for a product, they have a set of requirements for the product, features, and the opposing agent, for example, the reputation index, the threshold for cost-utility and time to live (TTL) for negotiation rounds. Once Alliance Engine finds the suitable party for an agent, it forwards a successful \texttt{matchAgent} message to the participating agents. 

\noindent \textbf{Negotiation Engine}: 
Negotiation Engine is the heart of the cloud marketplace system where the transactions among agents occur. Once the alliance is formed between two agents, a \texttt{commenceNegotiation} message with the timestamp in UTC is sent to the Negotiation Engines of the participant agents. Once the message is received by the agents, the rounds of offers and counter-offers continue between the agents until the cost or time to live (TTL) threshold values. Since the negotiation activities are performed via the Negotiation Engine, the cloud marketplace can also record the activity and perform analytical operations on the behavior of the buyer and seller agents. When Alliance Sentry notifies multiple agents that may be self-competing with each other, negotiation engines from each of those agents must report all negotiation transaction messages to the Master Negotiation Clearing House. 

\noindent \textbf{Behavior Watchdog}: The Behavior Watchdog service polls for the negotiation statistics and assigns each agent a behavior and reputation index. The reputation index is a rational number value \textit{R}, where 0 $\leq$ \textit{R} $\leq$ 1. More trustworthy and reputable agents possess higher value and tend to generate more agreements than lesser reputable agents. Additionally, it is essential for an agent to find out, in advance, an agent that is not only reputed but also ``well-behaved". Behavior Index, here, is also a rational number value \texttt{B}, where \texttt{B} $>$ 0, that showcases characteristic value about how an agent normally behaves during the negotiation \cite{nego_agent_hybrid_cloud}. The Behavior Index value can be important while designing the negotiation strategy.

\subsection{Cloud Agents}
Each BDI agent in the marketplace system has three database systems that store the agent's beliefs, goals, and plans. Beliefs are an agent's understanding about the current negotiation round for a feature or an issue \textit{i}. Beliefs are stored in the database named Beliefset and are constantly updated during the rounds of the negotiation. The goals are an agent's ruminative intentions to reach an agreement over a negotiation on a particular issue. With every round of negotiation, the agent constantly updates its goals to understand if the cost utility is within the thresholds for the next rounds of negotiation. For an agent, submitting a specific set of offers and counter-offers would be stored in the Plan Library. When an agent participates in the cloud marketplace, it assigns a weight of importance, \texttt{W}, to individual issues based on parameters like availability or dearth of a particular resource, where $\sum_{i=1}^{n} \texttt{W}_i = 1$. The negotiation agenda for an agent is decided based on who can fulfill the requirements of all the important issues for a product within the confines of the total cost utility that the agent can spend. Therefore, an agent stores every negotiation space it is participating in at any moment in Agenda DB. If resource availability for a service monitored by Resource Monitor crosses the threshold value set by the enterprise, the agent associated with that service may commence a negotiation with a seller agent in the marketplace to buy more resources. To enable multi-agent communication and negotiation from the same enterprise, the paper introduces two additional components as follows.

\noindent \textbf{Alliance Sentry}: The system must be prepared for the scenario where multiple agents from the same enterprise may compete with each other in the marketplace. For example, if agent A and agent B are negotiating with the seller agent X for a spot instance. As per the negotiation principle, the seller agent tries to maximize its cost utility function while the buyer agents try to minimize theirs. In this case, agent A and agent B will essentially compete with each other, even though they are from the same enterprise, increasing the overall cost for the enterprise. Therefore, Alliance Sentry acts as a self-hosted monitoring service that keeps track of all the alliances that have formed. Every \texttt{requestForQuote} message that is sent by the Alliance Engine service must pass through Alliance Sentry. If it detects an alliance of two  agents from the same enterprise with the same opposing agent, it sends a \texttt{cooperate} message with a communication channel ID to both the agents to advise them to co-ordinate and co-operate with each other to reduce the overall cost utility function for the enterprise.

\noindent \textbf{Master Negotiation Clearing House}: Once the Alliance Sentry detects a competing alliance, all involved agents must perform every subsequent negotiation round while coordinating with each other. The paper proposes a distributed message passing solution for the agents to communicate with each other. The Master Negotiation Clearing House acts as a ``mailbox" pool for distributed agents for message passing. The agents share every negotiation transaction message in subsequent rounds to the mailbox with the ID shared with all fellow competing agents. The section \ref{sec:negotiation} discusses the multi-agent communication and negotiation in detail.

\section{Multi-agent Communication and Negotiation}\label{sec:negotiation}
Before the start of the negotiation dialog with another agent, a BDI cloud agent submits its initial desire set \texttt{D} such as weight of an issue (\texttt{W\textsubscript{i}}), a range of acceptable cost value (\texttt{C\textsubscript{i}}) to each issue \texttt{i}, and a threshold value for the negotiation time period (\texttt{t\textsubscript{max}}) such that \texttt{t\textsubscript{min\textsubscript{i}}} $\leq$ \texttt{t\textsubscript{i}} $\leq$ \texttt{t\textsubscript{max\textsubscript{i}}}. \\
\indent Additionally, for an enterprise, the agents are assigned to individual services that negotiate with the sellers and buyers on the cloud marketplace. Since multiple agents belonging to the same enterprise can attempt to bid for the same resource with the attributes that is being sold by the same seller. This creates a conflicting situation where two or more agents from the same enterprise try to compete with each other for that resource. An unwanted side-effect of this scenario is that agents compete with each other thereby increasing the overall cost utility for an enterprise. In a multi-agent environment, agents may choose to perform following reactions viz., a) cooperate b) coordinate and c) negotiate. To reduce the overall cost utility for a product, this paper focuses on cooperation and coordination among agents. When a buyer agent seeks to establish an alliance with a suitable seller agent, it must submit an \texttt{RFQ} message to the cloud marketplace via the Alliance Engine. Alliance Sentry acts as a watchdog process that keeps track of all \texttt{RFQ} messages originating from agents in an enterprise. Alliance Sentry matches any requests from any agents from an enterprise that contain same parameters for a product and the seller agent. The Alliance Sentry registers such self-competing alliances to Alliance Registry. Once the alliances are confirmed by the participating agents, the Alliance Sentry will instruct the Negotiation Engine of every participating agent from an enterprise with a message, \texttt{\{agentId, productId, advertisementId, mailboxId\}}, with participating agent ID, the details about the product, details about the advertisement sent by the participating agent, and a mailbox ID. \\
\indent The paper proposes a message passing mechanism of mailboxes for agent communication. A mailbox provides an asynchronous FIFO message queue where a message appears in the queue as it is sent by an agent. A message can be blocking or non-blocking. Each mailbox has an unique ID and that mailbox can only be shared with selected participating agents. As a form of the message passing model, these mailbox connections must be set up prior to any communication. With every new negotiation round, an agent posts a message with its current offer, the utility of the offer and expectations for an agent from the negotiation round. Once an agent sends a message, the other participating agents treat this information about the current negotiation round as the new belief. Based on the new belief, the agents then calculate the goals for their offers and also construct new plans or modify the old ones. These steps are executed continuously until all other participating agents leave the self-competing alliances one by one. Apart from the communication, the agents also have to coordinate with each other to attempt to reach their goals while minimizing the overall cost incurred to the enterprise. As a result, the agents must be able to compromise their goals based on the beliefs that they learn from other agents. The paper introduces a cooperation threshold value \(\lambda\) for the agent behavior such that 0 $\leq$ \texttt{$\lambda$\textsubscript{i}} $\leq$ 1. This threshold value for agent behavior for multi-agent coordination depends on a) \texttt{c}, the change in the cost for an issue \texttt{i} in the current round compared to the previous rounds of negotiation. The cost delta is calculated as per the equation \ref{eq:cost_delta}. b) \texttt{t}, the time threshold value for an issue \texttt{i} where 0 $\leq$ \texttt{t\textsubscript{i}} $\leq$ 1 and it showcases the relative time remaining in the current negotiation and c) $\mu$, the resource urgency threshold for an issue \texttt{i} where 0 $\leq$ \texttt{$\mu$\textsubscript{i}} $\leq$ 1. The resource urgency threshold is calculated from the metrics fetched from the Resource Monitor service that constantly polls the infrastructure services to find the relative resource utilization for an issue that is being negotiated. Bigger value for this threshold indicates the higher urgency for the negotiation round to conclude.

\begin{equation}{\label{eq:cost_delta}}
  {c_i}_n = \frac{{c_i}_n - {c_i}_{n-1}}{{c_i}_n} \text{ where } n \ge 3
\end{equation}

After combining all three thresholds, the overall agent behavior cooperation threshold value is calculate via the equation \ref{eq:threshold}. This threshold value dictates the agent behavior while cooperating with other participating agents from the same enterprise that are negotiating with the same cloud marketplace agent on same product issues.

\begin{equation}{\label{eq:threshold}}
  \lambda_i = \sum_{i=1}^{n} c_i \times \mu_i \times t_i \text{  where } \sum_{i=1}^{n} c_i = 1, \sum_{i=1}^{n} \mu_i = 1, \sum_{i=1}^{n} t_i = 1
\end{equation}

\begin{figure}[t]
  \centering\includegraphics[width=0.7\linewidth]{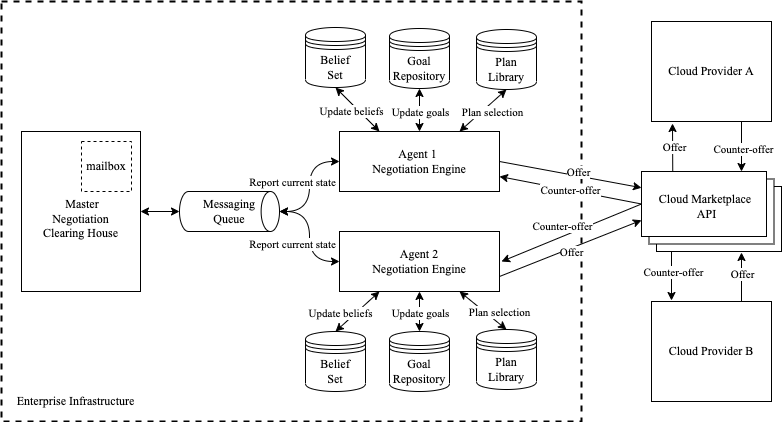}
  \caption{BDI Multi-agent Cooperation and Negotiation}
  \label{fig:negotiation}
\end{figure}

Every agent participating in the multi-agent cooperation sets the initial value for $\lambda$\textsubscript{$\phi$} for every issue \texttt{i} as a part of its belief set. As negotiation rounds progress for other agents, a new value is calculated for $\lambda$\textsubscript{i} and the new value is checked against the initial threshold value the agent is willing to commit to. The agent behavior for a negotiation round \texttt{N} is modified based on the cooperation threshold value as showcased in the equation \ref{eq:behavior_threshold}.

\begin{equation}\label{eq:behavior_threshold}
    \mathcal{N\textsubscript{i}} = 
    \begin{cases}
    \text{$\lambda$\textsubscript{i} $<$ $\lambda$\textsubscript{$\phi$}}, \text{where agent a tends to be headstrong}\\
    \text{$\lambda$\textsubscript{i} = $\lambda$\textsubscript{$\phi$}}, \text{where agent a tends to be linear}\\
    \text{$\lambda$\textsubscript{i} $>$ $\lambda$\textsubscript{$\phi$}}, \text{where agent a tends to be conceder}\\
    \end{cases}
\end{equation}

\indent When the negotiation rounds start between the participating agents from an enterprise and the cloud marketplace, there are multiple ways of how an agent records its response, \textit{R}. For a transaction between agents a and b, an agent may prepare the offer \texttt{O\textsubscript{i}\textsuperscript{t\textsubscript{n-1}}} for an issue \texttt{i}, if the total utility of the offer after the measurement of cooperation threshold value for the agent is greater than the offer received by the opponent. Otherwise, the agent sends a counter-offer to the opponent with a modified cost value for an issue \texttt{i}. Based on the prior agreements between two agents, if an counter-offer is received beyond the threshold value of the negotiation time period, the agent would terminate the negotiation round by sending \texttt{terminateNegotiation} message to the opponent. The offers and counter offers are generated based on an agent's own current beliefs as well as the shared beliefs of other self-competing agents from the enterprise.

\indent In the proposed model in this paper, the agents use a hybrid time and resource-dependent deadline-based tactics that use the behavior threshold value $\lambda$ to generate offers and counter-offers. As mentioned in \cite{nego_time_dep}, other similar tactics like time-dependent tactic, the agent behavior is governed by a time-based function \textit{f(t)} where an agent must try to pursue the closure of the negotiation within the threshold value of t\textsubscript{max}, where 0 $<$ t $\leq$ t\textsubscript{max}. In the hybrid time-resource dependent counter-offer tactic, a hardheaded tactic is applied near the negotiation threshold or if opponent's resources run short. When an agent practices hardheaded negotiation tactic, the agent attempts to lose very little utility for an issue and the concessions are made near the end of the time threshold, t\textsubscript{max} or resource threshold, \texttt{r\textsubscript{max}}, whichever is smaller. On the other hand, an agent practises a conceder tactic if it is desperate to reach a deal quickly by applying concessions to its offers starting early in the negotiation rounds. An agent may start conceding in the early rounds of the negotiation and then depending on the resource availability, it may choose to alter the tactic to become more hardheaded. If the agent finds that the opponent is practising a linear conceding tactic then the agent will try to match it, changing the counter-offers only slightly. Moreover, if the agent observes a conceding strategy by the opponent and the agent may choose to practise a little hardheaded tactic thereby attempting to improve the overall utility for an issue . On the other hand, if the counter-offers convey opponent's headstrong tactic, the agent chooses conceding tactic to facilitate faster agreement.

\section{Future Work}\label{sec:future_work}
The paper proposed assigning a BDI agent for every infrastructure service for an enterprise that chooses to autonomously buy or sell cloud resources based on the demand. The author believes that the idea proposed in the paper would be helpful in building an autonomously managed service offering, for example, as mentioned in \cite{bigbird}. While this facilitates the granular control over reducing the cost incurred during the negotiation phase, the paper only focuses on co-ordination and co-operation aspects of the multi-agent system. The author further intends to examine the third important aspect of the multi-agent system that is negotiation and study the feasibility of having self-competing agents negotiate with each other and decide who gets to negotiate with the opponent agent from the cloud marketplace first. Furthermore, another part of the future work would be to design a protocol to facilitate the multi-agent communication for an enterprise using methodologies like Law Governed Interaction \cite{lgi}.

\section{Conclusion}\label{sec:conclusion}
The paper showcased a BDI agents-based multi-agent system for cloud marketplace negotiation that enables multi-party, multi-issue negotiation for cloud resources. The idea proposed in the paper assigns a BDI agent to every infrastructure component that wishes to buy or sell cloud resources autonomously via the cloud marketplace system. To prevent agents competing with each other and driving the expenditure up for an enterprise, the paper presented a time-resource-based multi-agent co-operation and co-ordination tactic that would enable them choose an appropriate negotiation tactic with an opponent agent from the marketplace. Finally, the paper discusses the future improvements and additions like multi-agent negotiation for self-competing agents from an enterprise and designing a new protocol for multi-agent communication that is based on the predetermined set of rules.
%
%
%
%

\end{document}